\documentclass{scspaperproc}

\usepackage{latexsym}
\usepackage{graphicx}
\usepackage{mathptmx}

\usepackage{amsmath}
\usepackage{amsfonts}
\usepackage{amssymb}
\usepackage{amsbsy}
\usepackage{amsthm}

\usepackage[pdftex,colorlinks=true,urlcolor=blue,citecolor=black,anchorcolor=black,linkcolor=black,bookmarks=false,hidelinks]{hyperref}

\usepackage{hyphenat}
\newtheoremstyle{scsthe}
{8pt}
{8pt}
{\it}
{}
{\bf}
{.}
{.5em}
{}

\theoremstyle{scsthe}

\begin{document}

%
%

\pagestyle{fancyplain}

\thispagestyle{plain}
\firstPageHead{}

\chead{\fancyplain{}{\itshape\small Jayanetti, Frydenlund, and Weigle \vspace{8pt}}}

\rhead{}
\cfoot{}
\renewcommand{\headrulewidth}{0pt} 

\makeatletter
\let\@internalcite\cite
\def\cite{\def\@citeseppen{-1000}%
    \def\@cite##1##2{(##1\if@tempswa , ##2\fi)}%
    \def\citeauthoryear##1##2##3{##1 ##3}\@internalcite}
\def\citeNP{\def\@citeseppen{-1000}%
    \def\@cite##1##2{##1\if@tempswa , ##2\fi}%
    \def\citeauthoryear##1##2##3{##1 ##3}\@internalcite}
\def\citeN{\def\@citeseppen{-1000}%
    \def\@cite##1##2{##1\if@tempswa, ##2)\else{}\fi}%
    \def\citeauthoryear##1##2##3{##1 (##3)}\@citedata}
\def\citeA{\def\@citeseppen{-1000}%
    \def\@cite##1##2{(##1\if@tempswa , ##2\fi)}%
    \def\citeauthoryear##1##2##3{##1}\@internalcite}
\def\citeANP{\def\@citeseppen{-1000}%
    \def\@cite##1##2{##1\if@tempswa , ##2\fi}%
    \def\citeauthoryear##1##2##3{##1}\@internalcite}
\def\shortcite{\def\@citeseppen{-1000}%
    \def\@cite##1##2{(##1\if@tempswa , ##2\fi)}%
    \def\citeauthoryear##1##2##3{##2 ##3}\@internalcite}
\def\shortciteNP{\def\@citeseppen{-1000}%
    \def\@cite##1##2{##1\if@tempswa , ##2\fi}%
    \def\citeauthoryear##1##2##3{##2 ##3}\@internalcite}
\def\shortciteN{\def\@citeseppen{-1000}%
    \def\@cite##1##2{##1\if@tempswa, ##2\else{}\fi}%
    \def\citeauthoryear##1##2##3{##2 (##3)}\@citedata}
\def\shortciteA{\def\@citeseppen{-1000}%
    \def\@cite##1##2{(##1\if@tempswa , ##2\fi)}%
    \def\citeauthoryear##1##2##3{##2}\@internalcite}
\def\shortciteANP{\def\@citeseppen{-1000}%
    \def\@cite##1##2{##1\if@tempswa , ##2\fi}%
    \def\citeauthoryear##1##2##3{##2}\@internalcite}
\def\citeyear{\def\@citeseppen{-1000}%
    \def\@cite##1##2{(##1\if@tempswa , ##2\fi)}%
    \def\citeauthoryear##1##2##3{##3}\@citedata}
\def\citeyearNP{\def\@citeseppen{-1000}%
    \def\@cite##1##2{##1\if@tempswa , ##2\fi}%
    \def\citeauthoryear##1##2##3{##3}\@citedata}
%
%
%
\def\@citedata{%
    \@ifnextchar [{\@tempswatrue\@citedatax}%
                  {\@tempswafalse\@citedatax[]}%
}

\def\@citedatax[#1]#2{%
\if@filesw\immediate\write\@auxout{\string\citation{#2}}\fi%
  \def\@citea{}\@cite{\@for\@citeb:=#2\do%
    {\@citea\def\@citea{, }\@ifundefined
       {b@\@citeb}{{\bf ?}%
       \@warning{Citation `\@citeb' on page \thepage \space undefined}}%
{\csname b@\@citeb\endcsname}}}{#1}}%

%
\def\@citex[#1]#2{%
\if@filesw\immediate\write\@auxout{\string\citation{#2}}\fi%
  \def\@citea{}\@cite{\@for\@citeb:=#2\do%
    {\@citea\def\@citea{, }\@ifundefined
       {b@\@citeb}{{\bf ?}%
       \@warning{Citation `\@citeb' on page \thepage \space undefined}}%
{\csname b@\@citeb\endcsname}}}{#1}}%

%
\def\@biblabel#1{}
\makeatother

\newdimen\bibindent
\bibindent=.25in

\def\thebibliography#1{\section*{\refname}\list
   {}{\settowidth\labelwidth{[#1]}
   \leftmargin \bibindent
   \itemindent -\bibindent
   \listparindent \itemindent
	 \itemsep 4pt
   \parsep 0pt
   \usecounter{enumi}}
   \def\newblock{}
   \sloppy
   \sfcode`\.=1000\relax}

\setlength{\baselineskip}{12.7pt}

\def\SCSconferenceacro{MSVSCC'23}

\def\SCSpublicationyear{2023}

\def\SCSconferencedates{April 20}

\def\SCSconferencevenue{Suffolk, VA, USA}

\title{Exploring Xenophobic Events through GDELT Data Analysis}

\author{
Himarsha R. Jayanetti \\ 
Department of Computer Science \\
Old Dominion University \\
Norfolk, VA, USA \\
hjaya002@odu.edu\\
\and
Erika Frydenlund \\ 
Virginia Modeling, Analysis and Simulation Center \\
Old Dominion University \\
Suffolk, VA, USA \\
efrydenl@odu.edu\\
\and
Michele C. Weigle \\
Department of Computer Science \\
Old Dominion University \\
Norfolk, VA, USA \\
mweigle@cs.odu.edu\\
}

\maketitle

\section*{Abstract}

This study explores xenophobic events related to refugees and migration using the GDELT 2.0 database and APIs through visualizations. We conducted two case studies -- the first being an analysis of refugee-related news following the death of a two-year-old Syrian boy, Alan Kurdi, and the second a surge in news articles in March 2021 based on the data obtained from GDELT API. In addition to the two case studies, we present a discussion of our exploratory data analysis steps and the challenges encountered while working with GDELT data and its tools. 


\textbf{Keywords:}  Xenophobia, Refugees and Migrants, GDELT, Big Data, Data Science for Social Good

\section{Introduction}
\label{sec:intro}
People move around the world in pursuit of better opportunities or to flee conflicts and natural disasters. There are 281 million international migrants, or one in every 30 people worldwide~\cite{436}, and more than 82 million of them have been forcefully displaced~\cite{unhcrfaag}. These migrants make an effort to fit in with the host communities. However, widespread xenophobic and racist violence makes it difficult to uphold societal order and provide equal access to opportunities, resources, and even human dignity. Hence, it is imperative to study such xenophobic incidents and examine the underlying factors contributing to hostile behavior towards refugees in order to fight xenophobia. The United Nations High Commissioner for Refugees (UNHCR)\footnote{\url{https://www.unhcr.org/en-us/}} and the International Organization for Migration, (IOM)\footnote{\url{https://www.iom.int/}} which are responsible for promoting secure and well-organized migration, also have the responsibility to combat against xenophobia.

In our study, we use a massive and regularly updated dataset of online and TV and news reporting from GDELT\footnote{\url{https://www.gdeltproject.org/}} to explore xenophobic events. We leveraged Google BigQuery\footnote{\url{https://cloud.google.com/bigquery}} and GDELT APIs to extract and access large amounts of data of online and TV news coverage related to refugee and migration topics. BigQuery is a cloud-based data warehousing tool that enables us to query large datasets like GDELT quickly and efficiently. The GDELT API is an open API that provides us access to the massive GDELT database. By using both tools, we were able to extract valuable insights from the vast amounts of data available, allowing us to better understand the dynamics of xenophobic events and how they are portrayed in the media.

The objective of this paper is to describe how we conducted an exploratory data analysis phase, where we concentrated on particular case study events involving refugees and migration. These events were selected based on our prior research and expertise in the subject matter. Through our analysis of news article data, we were able to identify patterns surrounding specific events by examining the time periods before and after the occurrence of the event. By focusing on these case study events, we were able to see how the media coverage surrounding the topics of refugees, migration, and xenophobia significantly increased around the time of the event. We gained a deeper understanding of how the media portrays these issues using different factors such as the type of event, the overall tone of the news media coverage, and the different actors and countries involved. These insights allow us to better identify and analyze patterns and themes in the data, which can help inform future research in developing better and more suitable interactive visualizations that monitor xenophobic violence against refugees and migrants.

\section{Background and Related Work}
\label{sec:relwork}

In this study, we used Global Data on Events, Location, and Tone (GDELT), which is a digital news database of geolocated events worldwide from 1979 to the present. GDELT, which has billions of records and is continuously updated in real time, is a prime example of big data. The GDELT databases use the Conflict and Mediation Event Observations (CAMEO) taxonomy, a framework for coding event-related data, to automatically code data for use in research~\cite{gerner2002conflict}. 

Researchers in the past have used GDELT data for a variety of studies, such as studying the effects of civil unrest, complexity in terms of political activities, and capturing peace through the Global Peace Index (GPI)~\cite{yonamine2013nuanced,fang2016identifying,voukelatou2022understanding}. Vargo et al. studied the power of fake news from 2014 to 2016 in online news media using the GDELT dataset~\cite{vargo2018agenda}. Their research revealed that although the prevalence of fake news has risen, these websites do not possess undue influence. Various researchers also used social media platforms (like Twitter\footnote{\url{https://twitter.com/}}) as well as newspaper articles in opinion mining  about a range of topics from online education to industrial production~\cite{fu2020opinion,tilly2021macroeconomic}. 

Various studies have examined forced migration and policy implications in countries that support migration~\cite{frydenlund2019mobility,frydenlund2022opportunities,frydenlund2019characterizing}. Yesilbas et al. utilized GDELT to build a large dataset of global news to study the tone, volume, and topics of media coverage of refugees~\cite{yesilbas2021analysis}. They uncovered that the reason for the negative tone was both because of the anti-migrant sentiment as well as sorrow and empathy for the refugees. Although our research shares similarities with previous studies in that we aim to analyze events related to refugee and migrant communities, as well as the sentiment towards them, our approach differs from previous analyses and datasets. Specifically, our objective is to develop a monitoring system in real-time for xenophobic events utilizing GDELT data to identify potential hotspots of violence, thereby enabling us to predict and prevent any escalation.

While combating xenophobia is within the mandates of international organizations such as UNHCR and IOM, there is no worldwide tool for tracking these events. The Internal Displacement Monitoring Centre\footnote{\url{https://www.internal-displacement.org}} has designed a hand-coded data synthesis tool to monitor migration caused by natural disasters. This tool has been widely adopted by major humanitarian organizations.  ACLED\footnote{\url{https://acleddata.com/}} similarly tracks protest and violence across the world. Xenowatch\footnote{\url{https://www.xenowatch.ac.za}} is an online heatmap using data that researchers have hand-coded from user-submitted news articles of xenophobic events in South Africa. These tools have had immeasurable impacts on research and policy-level decision-making, but no such tool exists on the global scale for xenophobic events and actions against migrants. The ultimate aim of this study is to construct such a tool; however, this paper will solely discuss our preliminary exploratory analysis.



\section{Methodology}
\label{sec:method}

In this section, we will provide an overview of the data utilized in the study, explain our data collection process, present the findings of several case studies carried out including visualizations, and lastly, discuss the challenges encountered at different stages of this preliminary research. While the long-term goal of this project is to develop a monitoring dashboard for xenophobic violence worldwide, at this point we are in the exploratory data analysis and dashboard design phase. 

\subsection{Data}
\label{subsec:data}

In our study, we are using the GDELT 2.0 database, which is updated every 15 minutes and translates articles from around the world from 65 different languages into English~\cite{gdelt2}.

\subsubsection{Understanding the Data}
\label{subsubsec:understanding_data}

We have dedicated a substantial amount of time on this project to fully understand the data that is available to us. Through our efforts, we were able to identify three key tables in the database that hold essential data for our analysis of xenophobia using GDELT data.

\begin{enumerate}
    \item \textbf{Event}: This table contains data about events happening globally. Each row in the Event Table represents a single event. Each event is coded with information such as an event identification number (\texttt{GLOBALEVENTID}), actors (\texttt{Actor1Code}, \texttt{Actor2Code}, action, and location. \\
    
    \item \textbf{Event Mentions}: This table contains a row for each mention of the event in a news article or other source. Each mention is coded with its respective \texttt{GLOBALEVENTID} (which allows linking to the Event Table) and information about the tone of the mention (positive or negative). Specifically, each mention row in Event Mentions contains a \texttt{GLOBALEVENTID} that corresponds to the \texttt{GLOBALEVENTID} of the event that it mentions in the Event Table. This table contains an external identifier (\texttt{MentionIdentifier}) for the source document, which can be utilized to uniquely identify the document.   \\
    
    \item \textbf{Global Knowledge Graph (GKG)}~\cite{gdeltgkg}: This table connects data from various sources to form an extensive interconnected network that encapsulates everything including events around the world, their corresponding contexts, associated actors, and the overall sentiment of media coverage surrounding the event. The \texttt{DocumentIdentifier} field in the table corresponds to the \texttt{MentionIdentifier} in the Event Mentions table. Additionally, this table comprises the results obtained from the Global Content Analysis Measures (GCAM) system, which employs multiple cutting-edge content analysis tools to capture over 2,230 latent dimensions for each news article monitored by GDELT~\cite{gdeltgcam}.
    
\end{enumerate}

\begin{figure}[htb]
{
\centering
\includegraphics[width=0.9\textwidth]{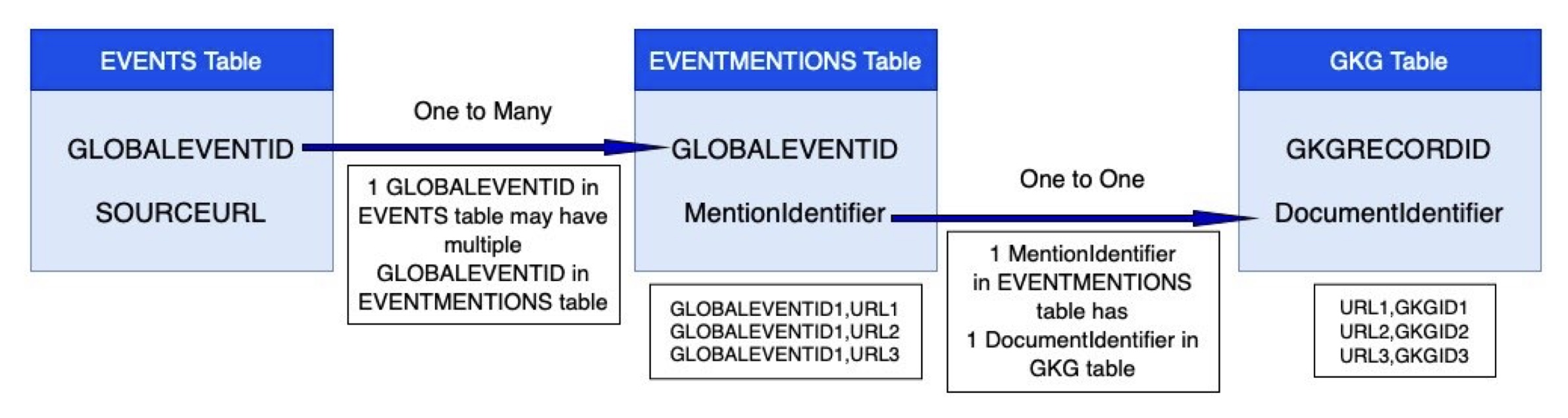}
\caption{The relationship among the three primary database tables: Events, Event Mentions, and GKG}\label{fig:DBtable}
}
\end{figure}

We have illustrated the database schema of the above three tables in Figure~\ref{fig:DBtable} to the best of our understanding thus far. As shown in Figure~\ref{fig:DBtable}, there exists a one-to-many relationship between the \texttt{GLOBALEVENTID} fields in the Event table and the Event Mentions table. This is due to the fact that each mention of an event corresponds to a row in the Mentions table. The \texttt{MentionIdentifier} field in the Mentions table can be utilized to merge the Mentions table with the GKG table. Our understanding of the connections between these tables facilitated our ability to aggregate, filter, and merge database tables as necessary to obtain the desired output.


\subsubsection{Data Collection Methods and Criteria}
\label{subsubsec:data_collection}

The GDELT database is a large, open-source database and is supported by Google in the form of cloud computing resources that help users access the data using BigQuery, which uses SQL-like queries. We used Google BigQuery to analyze the data as it can handle large amounts of data without requiring any additional setup or configuration.

\begin{figure}[htb]
{
\centering
\includegraphics[width=0.9\textwidth]{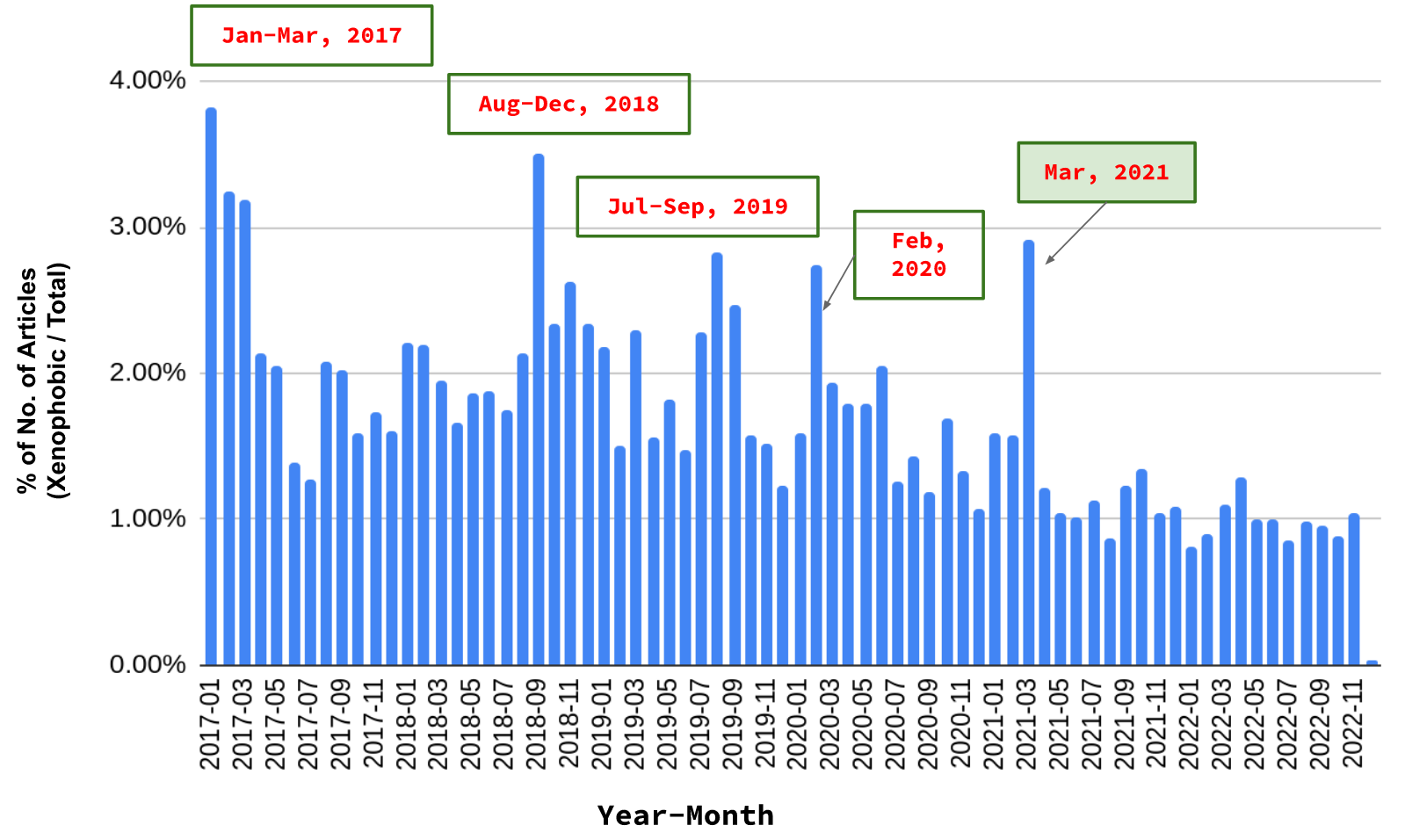}
\caption{The number of articles with xenophobic themes as a percentage of total article count over time (from January 2017 to December 2022). This data was obtained using the GDELT API.}\label{fig:gdeltAPIout}
}  
\end{figure}

We identified two distinct criteria to collect data for a specific time frame. We used these two criteria to obtain two separate datasets for our case studies as discussed in Section~\ref{subsec:analysis}. 

 \begin{enumerate}
    \item Criteria 1: Events where \texttt{Actor2Code} is \texttt{REF} (indicative of Actor 1 performing the act on ‘refugee’). \\

    \item Criteria 2: Events where \texttt{Actor2Code} is \texttt{REF} and has GKG themes that pertain to refugees. \\
    
\end{enumerate}

As our first dataset, we queried the data satisfying criteria 1 where the \texttt{Actor2Code} was \texttt{REF} by connecting to Google Big Query through a Jupyter Notebook.  We were able to download the data in a comma-separated values file (CSV). This will be further explained in Section~\ref{subsubsec:alan_kurdi}. 

As our second dataset, we queried the data that met criteria 2. We identified eight main GKG themes that are relevant to refugees:
    
    \begin{itemize}
        \item \texttt{DISCRIMINATION\_IMMIGRATION\_XENOPHOBIA}
        \item \texttt{DISCRIMINATION\_IMMIGRATION\_ANTIIMMIGRANTS}
        \item \texttt{DISCRIMINATION\_IMMIGRATION\_OPPOSED\_TO\_IMMIGRANTS}
        \item \texttt{DISCRIMINATION\_IMMIGRATION\_AGAINST\_IMMIGRANTS}
        \item \texttt{DISCRIMINATION\_IMMIGRATION\_ATTACKS\_ON\_IMMIGRANTS}
        \item \texttt{DISCRIMINATION\_IMMIGRATION\_ATTACKS\_AGAINST\_IMMIGRANTS}
        \item \texttt{DISCRIMINATION\_IMMIGRATION\_XENOPHOBE}
        \item \texttt{DISCRIMINATION\_IMMIGRATION\_XENOPHOBES} \\
    \end{itemize}
    
From now on, we will use the term ``GKGthemes\_REF'' to refer to these eight themes. We made use of the GDELT API\footnote{\url{https://blog.gdeltproject.org/gdelt-doc-2-0-api-debuts/}} to gain insight into the data before we query for the data itself. We used the GDELT 2.0 Doc API Client,\footnote{\url{https://github.com/alex9smith/gdelt-doc-api}} a Python client to fetch data from the GDELT API. By using the \texttt{timelinevolraw} option in this Python library, we obtained the number of articles matching the theme filter and the total news articles monitored by GDELT over time. \\
    
Figure~\ref{fig:gdeltAPIout} illustrates the variation of the number of articles with xenophobic themes as a percentage of the total article count over time (monthly data from January 2017 to December 2022). We noticed several spikes in  the chart, which indicate a surge in the number of articles compared to the other months. For our second case study, we decided to download the data from March 2021 (spike highlighted in green in Figure~\ref{fig:gdeltAPIout}), where \texttt{Actor2Code} is \texttt{REF} and is categorized by GDELT as a ``GKGthemes\_REF'' (\texttt{V2Themes} like \texttt{DISCRIMINATION\_IMMIGRATION}). This will be further discussed in Section~\ref{subsubsec:spa_shooting}.

\subsection{Data Download and Analysis Results}
\label{subsec:analysis}

In this exploratory data analysis phase, we focused on certain case study events that we knew from earlier research and subject matter expertise that caused significant coverage of refugees, migration, and xenophobic sentiments. Our methodology involved first selecting a specific incident related to refugees that received significant attention. We then developed a hypothesis that we aim to test using data analysis. We downloaded the data as described above from the GDELT database around the chosen incident, covering a period ranging from a few months before the event to a few months after. Next, we conducted a feature identification process, in which we identified the characteristics that influence or contribute to the observed outcomes. Finally, we compared the findings from the analysis with our original hypothesis to assess whether our hypothesis was confirmed or disproved. This approach enabled us to gain a deeper understanding of the factors that impact refugee-related issues.

\subsubsection{Case Study 1: Alan Kurdi Incident}
\label{subsubsec:alan_kurdi}
Our first case study was set around the death of a two-year-old Syrian boy, Alan Kurdi, born as Alan Shenu and initially reported as Aylan Kurdi~\cite{125sentence}. In September 2015, Alan and his family, who were refugees from Syria, attempted to travel to Europe from Turkey. Tragically, Alan, along with his mother and brother lost their life by drowning in the Mediterranean Sea while undertaking this perilous journey. The \href{https://static01.nyt.com/images/2020/03/13/world/13Turkey-migrant-boy/merlin_102645733_13c2554d-47cc-4802-9d98-145601d26e33-superJumbo.jpg?quality=75&auto=webp}{photograph of Alan's body lying face down on a Turkish beach} brought the incident to the forefront of international attention.

\begin{figure}[htb]
{
\centering
\includegraphics[width=1\textwidth]{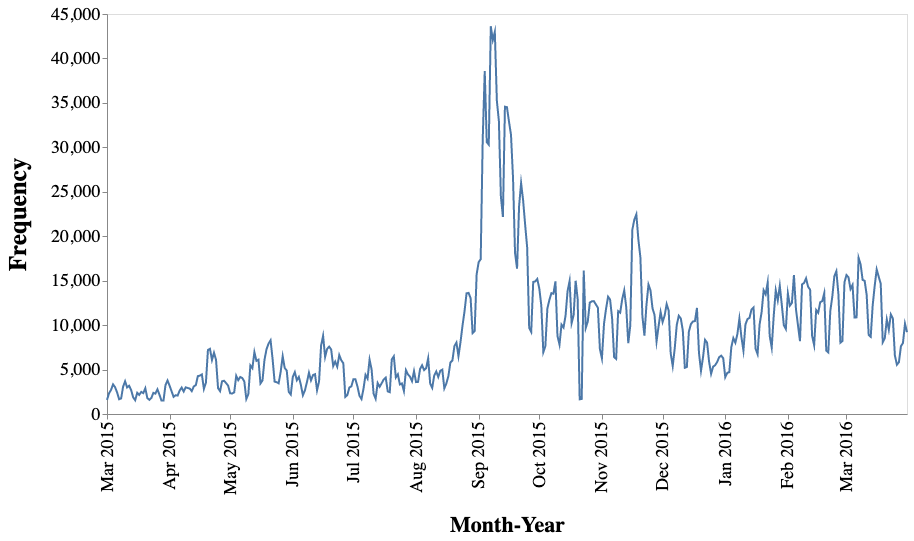}
\caption{The number of news articles from March 2015 to March 2016 where \texttt{Actor2Code} is \texttt{REF}.}
\label{fig:cs1numArticles}
}
\end{figure}

We downloaded the data six months before and after the incident (March 2015 to March 2016) where \texttt{Actor2Code} is \texttt{REF}.\footnote{\url{https://github.com/himarshaj/GDELT_ExploratoryAnalysis_XenophobicEvents/blob/main/Data/AK_before_after.zip}} Figure~\ref{fig:cs1numArticles} illustrates the timeline of the number of news articles, which confirmed a significant surge in attention to refugee-related news around reports of Alan Kurdi's death.

\begin{figure}[htb]
    \centering
    \begin{minipage}{0.49\textwidth}
        \centering
        \includegraphics[width=1\textwidth]{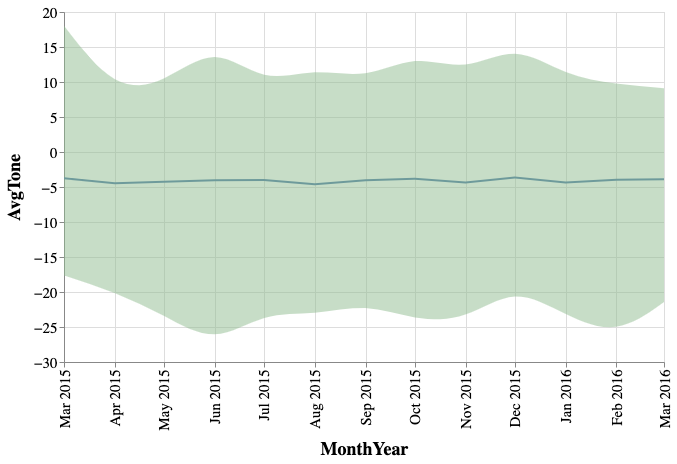} 
        \caption{\texttt{AvgTone} from March 2015 to March 2016 (line is median, area shows range)}
        \label{fig:avgtone}
    \end{minipage}\hfill
    \begin{minipage}{0.49\textwidth}
        \centering
        \includegraphics[width=1\textwidth]{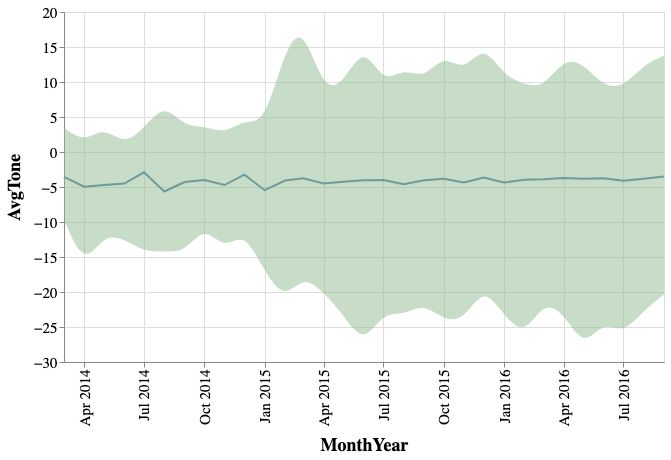} 
        \caption{\texttt{AvgTone} from March 2014 to March 2016 (line is median, area shows range)}
        \label{fig:avgtone_extended}
    \end{minipage}
\end{figure}

We also examined the \texttt{AvgTone} of the news articles around the time to understand the sentiment of news articles. Figure~\ref{fig:avgtone} shows the variation of  \texttt{AvgTone} over time in an area and line chart. The line in blue shows the \texttt{median} value whereas the green area covers the \texttt{min} and \texttt{max} values for \texttt{AvgTone} over time. Figure~\ref{fig:avgtone} shows that the sentiment of the news articles remained consistently negative over time without any abrupt changes in \texttt{AvgTone}. To explore further, we extended the timeline further back in time (to March 2014) and visualized the variation of \texttt{AvgTone} over time, which is presented in Figure~\ref{fig:avgtone_extended}. Despite the \texttt{median} of \texttt{AvgTone} still remaining consistent on the negative side, we noticed a shift in the range (a higher gap between \texttt{min} and \texttt{max}) before and after the start of January 2015.

\subsubsection{Case Study 2: Spike in Number of Articles around March 2021}
\label{subsubsec:spa_shooting}

\begin{figure}[htb]
    \centering
    \begin{minipage}{0.49\textwidth}
        \centering
            \includegraphics[height=0.6\textwidth]{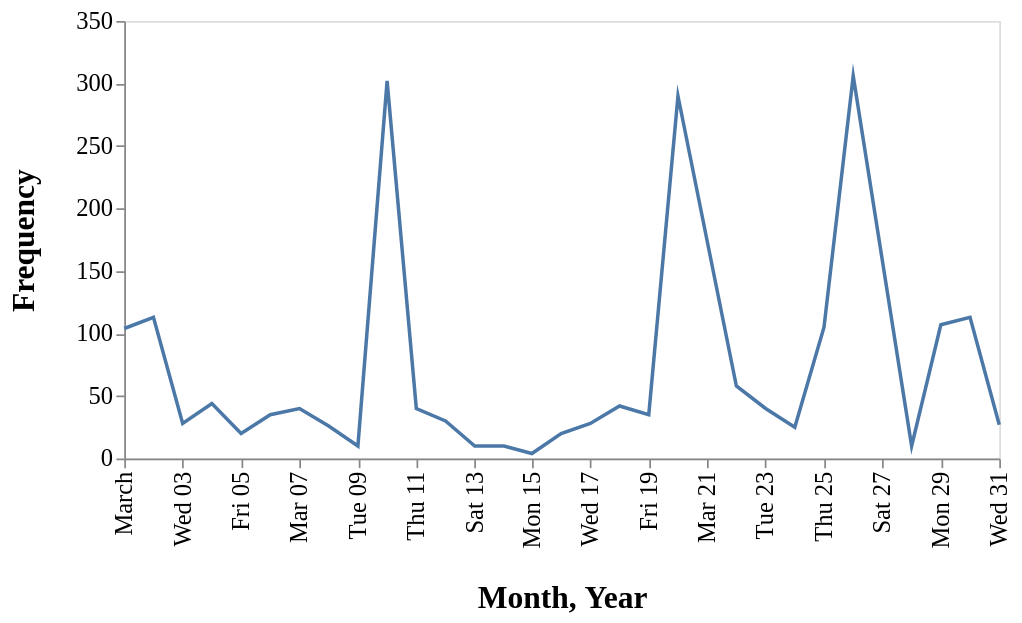}
            \caption{Variation in the number of articles during March 2021}
            \label{fig:cs2_NumArticles}
    \end{minipage}\hfill
    \begin{minipage}{0.49\textwidth}
        \centering
        \includegraphics[height=0.6\textwidth]{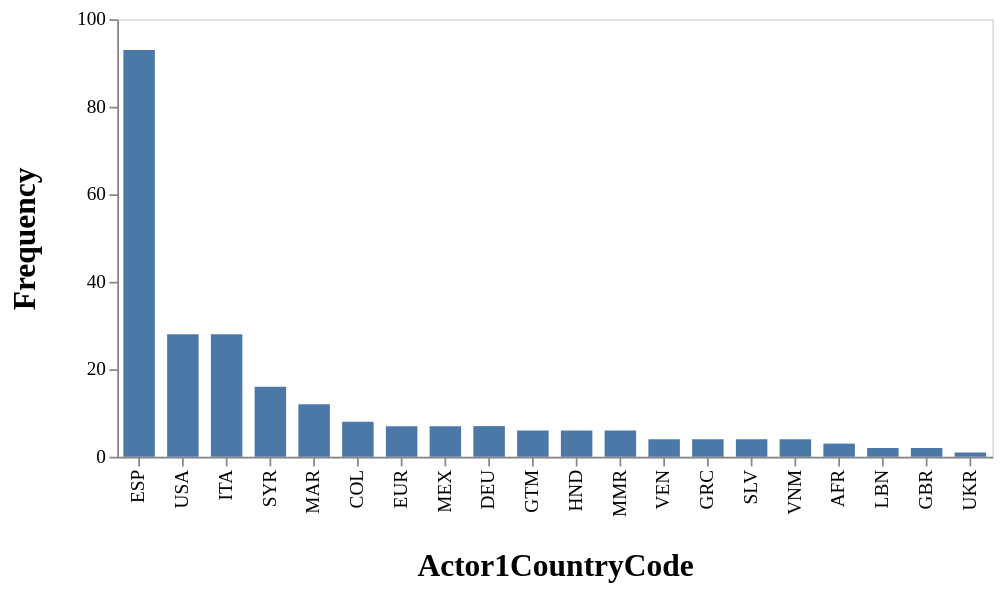}
        \caption{Frequency of the top 20 most prevalent \texttt{Actor1CountryCode}}
        \label{fig:cs_2Actor1CountryCode}
    \end{minipage}
\end{figure}

In this case study, we focused on investigating the surge in the number of articles in March 2021, which we observed through the GDELT API, as described in Section~\ref{subsubsec:data_collection}. Our initial hypothesis was that this surge is due to the  shootings that targeted three separate spas  in Atlanta, Georgia on March 16, 2021~\cite{8dead}. The fact that the majority of the victims of the shootings were women of Asian descent led to widespread outrage. This incident brought attention to the concerning increase in hate crimes and discrimination aimed at Asian communities in the United States. We downloaded the data for March 2021 where \texttt{Actor2Code} is \texttt{REF} and the theme was in the ``GKGthemes\_REF'' set.\footnote{\url{https://github.com/himarshaj/GDELT_ExploratoryAnalysis_XenophobicEvents/blob/main/Data/Actor2_REF_mar2021.csv}}  Figure~\ref{fig:cs2_NumArticles} shows the variation in the number of news articles during the month. It was surprising to observe that none of these peaks coincided with the date of the shooting incident.

\begin{figure}[htb]
{
\centering
\includegraphics[width=1\textwidth]{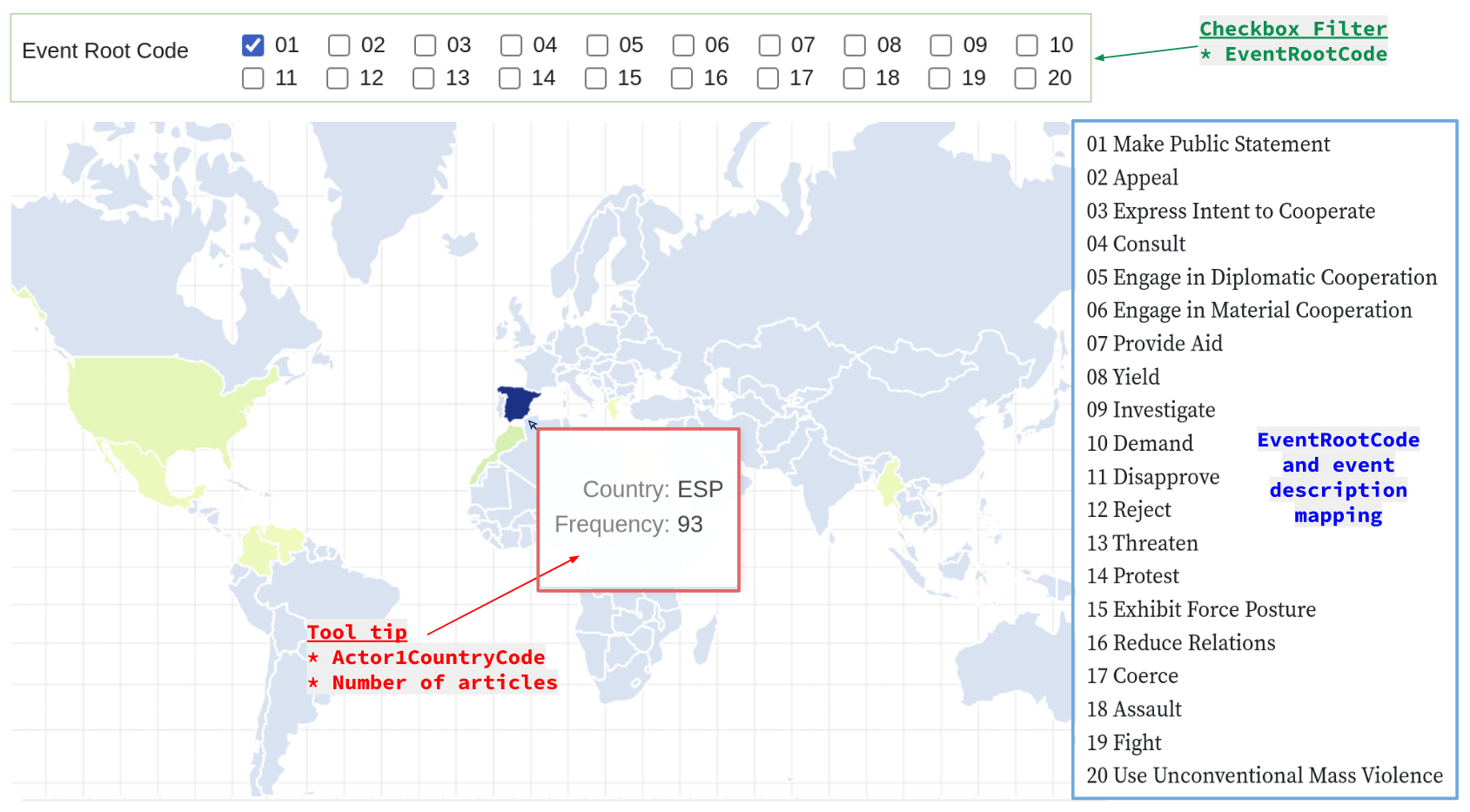}
\caption{A choropleth map  to visualize the number of articles based on location with a tooltip (highlighted in red) that displays the country and the corresponding frequency of articles. A checkbox filter (highlighted in green) for \texttt{EventRootCode} along with its description (highlighted in blue).}
\label{fig:map}
}
\end{figure}

To gain insight into the distribution of countries involved (via country code of Actor 1), we examined the frequency of the top 20 most prevalent \texttt{Actor1CountryCode} as shown in Figure~\ref{fig:cs_2Actor1CountryCode}. We observed that the highest frequency of articles based on \texttt{Actor2CountryCode} was \texttt{ESP} (the three-digit CAMEO code for Spain), followed by \texttt{USA} and then \texttt{ITA}. We used a choropleth map (Figure~\ref{fig:map}) to visualize the location-based data more effectively, providing a user-friendly perception. We incorporated a tooltip that displays the country and the corresponding frequency of articles into the map. We also included a checkbox filter for \texttt{EventRootCode} along with its description to increase interactivity and uncover more patterns in the data. For example, in Figure~\ref{fig:map} the checkbox filter was turned on for \texttt{EventRootCode} of value \texttt{01} which refers to ``Make Public Statement''. We used the machine-readable CAMEO event code that GDELT has made available alongside human-friendly event descriptions~\cite{cameodescr}. Upon examining the raw data, we were able to determine that the high number of articles with \texttt{Actor2CountryCode} as \texttt{ESP} was due to a significant number of articles reporting on the increase of African migrants arriving on the Canary Islands, which is an autonomous community of Spain, reported around March 26, 2021. The data analysis revealed that the last spike in the number of articles shown in Figure~\ref{fig:cs2_NumArticles} was linked to this event.

\subsection{Discussion}
\label{subsec:challenges}
In this section, we will discuss the lessons we have learned thus far and the challenges we have encountered during the course of this study. One of the significant challenges we encountered was in comprehending the available data, even though there is documentation publicly available. Some of the reasons that contribute to the complexity of understanding the data and how we attempted to overcome those challenges are outlined below.

 \begin{enumerate}
    \item The complexity of the database and the vast amount of data it contains made it very challenging for us to get started with the study. As discussed in Section~\ref{subsubsec:understanding_data}, we spent a significant amount of time understanding the different tables and how they connect with each other. This proved to be beneficial in enabling us to proceed with our research more efficiently. \\
    
    \item A large number of columns in each database table made it difficult to understand what fields were important to consider. We spent time understanding each column in the database tables. Given our primary objective of identifying the factors that can be utilized to detect instances of xenophobic violence, all columns in the tables are crucial to us. Therefore, we refrained from excluding any columns in our queries, except in instances where multiple versions of the same data were available, in which case we selected only the most recent version.  \\
    
    \item Data quality issues arise since the database utilizes data from multiple sources. When scrutinizing the data, we identified that some of the fields may contain null or incorrectly labeled values, thereby introducing noise. We have decided to acknowledge such noise and recognize that it is an inherent aspect of large-scale data analysis. By focusing on the larger picture and analyzing the data as a whole, we can identify meaningful patterns and insights that may not be evident from individual entries in database.
 
 \end{enumerate}



During our study, we encountered a few challenges while working with certain tools. As a first-time user of BigQuery, we faced some complexities in setting up and utilizing the platform effectively. While it offers powerful data analysis capabilities, the initial learning curve was steep, and we had to spend a significant amount of time familiarizing ourselves with the interface and features. We also faced limitations while using the GDELT API. One such limitation was that the API only provides data on articles produced in 2017 and later, which somewhat restricted the scope of our initial analysis using the API. However, we made use of the GDELT API to some extent to gain insights about the data after 2017, as discussed in Section~\ref{subsubsec:data_collection}.


During the exploration of the visualizations, we faced several challenges. We iterated through various approaches to visualize the data and continue to refine our options based on the data types. We explored multiple options and approaches in analyzing the data, but have had to revise our methods or start over as we gained a better understanding of the data. As we identified the limitations of the data, we consistently addressed those limitations to refocus our efforts on the most relevant questions. We realized that there is no universal visualization approach that works for all types of data. As it is crucial to ensure the accuracy and validity of the analysis, we had to re-evaluate our approaches to accomplish this objective.

Despite the challenges that we faced, which sometimes required additional time and effort, they ultimately led us to gain a deeper understanding of the data and its limitations. This understanding allowed us to make better design choices and ask more precise research questions for future investigations.

\section{Future Work}
\label{sec:future}
Moving forward, our study's results can be used to develop better and more suitable interactive visualizations that monitor xenophobic violence against refugees and migrants. Building upon the insights gained from this study, our next steps involve developing impactful visualizations that can aid us in addressing the following questions:

\begin{enumerate}
    \item What countries/regions are the ``hotspots'' of xenophobia? We would like to explore the severity of these events based on the number of refugees and migrants living in the area.

    \item How to know when xenophobic outbreaks are escalating and to prioritize them before they reach a critical stage?

    \item Can we identify the underlying events that trigger an upswing in xenophobic violence?
\end{enumerate}

Our long-term goal is to expand the scope of our data visualization research beyond xenophobia and investigate its potential use in monitoring a broad spectrum of societal issues, including racism, global health disparities, and incidents of loss of life or property caused by climate change worldwide. These potential future applications could significantly enhance the understanding of these complex issues and aid in the development of effective solutions for those issues.

\section{Conclusion}
\label{sec:conclusion}

Migration is a common occurrence worldwide as individuals move to new countries to seek better opportunities or to escape natural disasters and conflicts. However, xenophobic and racist violence poses a challenge for migrants as it hinders them from blending into new communities. To combat xenophobia, it is important to examine and understand xenophobic events and underlying factors contributing to hostility towards refugees. We used the GDELT 2.0 database which includes online and TV news to explore xenophobic events. We made use of the BigQuery and GDELT APIs to efficiently access extensive GDELT data and conduct exploratory data analysis, with a focus on case study events related to refugees and migration. 

For our initial case study, we examined the period surrounding the passing of Alan Kurdi, a two-year-old Syrian boy who died in the Mediterranean Sea while attempting to migrate from Turkey to Europe with his family. We studied the amount of news coverage on refugees and the sentiment expressed in news articles both before and after the event. We found that there was a significant increase in media attention to refugee-related news after Alan Kurdi's death, and the sentiment remained mostly negative with no abrupt changes in the average tone. However, there was a shift in the range of emotions expressed before and after January 2015. Our second case study aimed to investigate a surge in the number of news articles in March 2021 that we found through our insights using the GDELT API. Our findings indicated that the spike in news articles was linked to the increase of African migrants arriving on the Canary Islands, with Spain being the most prominent country code. We used a choropleth map to visualize the data obtained from the location-based analysis. 

Throughout the study, we encountered various challenges, including the intricate nature of the database, data quality concerns, and the long learning curve of working with specific tools such as BigQuery, as well as the constraints of the GDELT API. Moreover, the visualization process presented its own unique set of challenges, necessitating several approaches and revisions. Despite these challenges, our exploratory study allowed for a better understanding of the data and develop various charts to effectively utilize and visualize data concerning different events.

\section*{Acknowledgments}
\label{sec:acknowledgments}
This research was funded under the project "Data Science for Social Good: Mining and Visualizing Worldwide News to Monitor Xenophobic Violence", through the 2022-2023 ODU Data Science Seed Funding Program. 



\bibliographystyle{scsproc}
\bibliography{refs}

\begin{thebibliography}{}

\bibitem[\protect\citeauthoryear{Fang, Gao, Fan, and Yang}{Fang
  et~al.}{2016}]{fang2016identifying}
Fang, P., J.~Gao, F.~Fan, and L.~Yang. 2016.
\newblock ``Identifying Political “hot” Spots Through Massive Media Data
  Analysis''.
\newblock In {\em Social, Cultural, and Behavioral Modeling: 9th International
  Conference, SBP-BRiMS 2016, Washington, DC, USA, June 28-July 1, 2016,
  Proceedings 9}, pp.  282--290.
\newblock Springer.

\bibitem[\protect\citeauthoryear{Frydenlund, Jones, and Padilla}{Frydenlund
  et~al.}{2019}]{frydenlund2019mobility}
Frydenlund, E., E.~C. Jones, and J.~J. Padilla. 2019.
\newblock ``Mobility in crisis: An agent-based model of refugees’ flight to
  safety''.
\newblock {\em Human Simulation: Perspectives, Insights, and Applications\/},
  pp. 191--208.


\bibitem[\protect\citeauthoryear{Frydenlund and Padilla}{Frydenlund and
  Padilla}{2022}]{frydenlund2022opportunities}
Frydenlund, E., and J.~J. Padilla. 2022.
\newblock ``Opportunities and Challenges of Using Computer-Based Simulation in
  Migration and Displacement Research: A Focus on Lesbos, Greece''.
\newblock {\em McGill-Queen’s Refugee and Forced Migration Studies Series
  editors: Megan Bradley and James Milner\/}, pp. 279.


\bibitem[\protect\citeauthoryear{Frydenlund, Yilmaz~{\c{S}}ener, Gore,
  Boshuijzen-van Burken, Bozdag, and De~Kock}{Frydenlund
  et~al.}{2019}]{frydenlund2019characterizing}
Frydenlund, E., M.~Yilmaz~{\c{S}}ener, R.~Gore, C.~Boshuijzen-van Burken,
  E.~Bozdag, and C.~De~Kock. 2019.
\newblock ``Characterizing the mobile phone use patterns of refugee-hosting
  provinces in turkey''.
\newblock {\em Guide to Mobile Data Analytics in Refugee Scenarios: The'Data
  for Refugees Challenge'Study\/}, pp. 417--431.


\bibitem[\protect\citeauthoryear{Fu, Yan, Meng, Wang, Hu, Li, Wang, He, and
  Wang}{Fu et~al.}{2020}]{fu2020opinion}
Fu, Z., M.~Yan, C.~Meng, W.~Wang, X.~Hu, Y.~Li, J.~Wang, Z.~He, and Z.~Wang.
  2020.
\newblock ``Opinion mining about online education basing on GDELT and Twitter
  data''.
\newblock In {\em 2020 IEEE/WIC/ACM International Joint Conference on Web
  Intelligence and Intelligent Agent Technology (WI-IAT)}, pp.  741--745.
\newblock IEEE.

\bibitem[\protect\citeauthoryear{Gerner, Schrodt, Yilmaz, and Abu-Jabr}{Gerner
  et~al.}{2002}]{gerner2002conflict}
Gerner, D.~J., P.~A. Schrodt, O.~Yilmaz, and R.~Abu-Jabr. 2002.
\newblock ``Conflict and mediation event observations (CAMEO): A new event data
  framework for the analysis of foreign policy interactions''.
\newblock {\em International Studies Association, New Orleans\/}.


\bibitem[\protect\citeauthoryear{{International Organization for
  Migration}}{{International Organization for Migration}}{2021}]{436}
{International Organization for Migration} 2021.
\newblock {\em World Migration Report 2022}.


\bibitem[\protect\citeauthoryear{{The GDELT Project}}{{The GDELT
  Project}}{}]{cameodescr}
{The GDELT Project}.
\newblock ``{CAMEO Eventcode and Description}''.
\newblock \url{https://www.gdeltproject.org/data/lookups/CAMEO.eventcodes.txt}.

\bibitem[\protect\citeauthoryear{{The GDELT Project}}{{The GDELT
  Project}}{2015}]{gdelt2}
{The GDELT Project} 2015.
\newblock ``{GDELT 2.0: Our Global World in Realtime}''.
\newblock
  \url{https://blog.gdeltproject.org/gdelt-2-0-our-global-world-in-realtime/}.

\bibitem[\protect\citeauthoryear{{The GDELT Project, GCAM}}{{The GDELT Project,
  GCAM}}{2014}]{gdeltgcam}
{The GDELT Project, GCAM} 2014.
\newblock ``{Introducing the Global Content Analysis Measures (GCAM)}''.
\newblock
  \url{https://blog.gdeltproject.org/introducing-the-global-content-analysis-measures-gcam/}.

\bibitem[\protect\citeauthoryear{{The GDELT Project, GKG}}{{The GDELT Project,
  GKG}}{2014}]{gdeltgkg}
{The GDELT Project, GKG} 2014.
\newblock ``{Introducing GKG 2.0 – The Next Generation of the GDELT Global
  Knowledge Graph}''.
\newblock
  \url{https://blog.gdeltproject.org/introducing-gkg-2-0-the-next-generation-of-the-gdelt-global-knowledge-graph/}.

\bibitem[\protect\citeauthoryear{{The New York Times}}{{The New York
  Times}}{2020}]{125sentence}
{The New York Times} 2020.
\newblock ``{3 Men Sentenced to 125 Years Each in Drowning of Syrian Refugee
  Boy}''.
\newblock
  \url{https://www.nytimes.com/2020/03/13/world/middleeast/alan-kurdi-death-trial.html}.

\bibitem[\protect\citeauthoryear{{The New York Times}}{{The New York
  Times}}{2021}]{8dead}
{The New York Times} 2021.
\newblock ``{8 Dead in Atlanta Spa Shootings, With Fears of Anti-Asian Bias}''.
\newblock
  \url{https://www.nytimes.com/live/2021/03/17/us/shooting-atlanta-acworth}.

\bibitem[\protect\citeauthoryear{Tilly, Ebner, and Livan}{Tilly
  et~al.}{2021}]{tilly2021macroeconomic}
Tilly, S., M.~Ebner, and G.~Livan. 2021.
\newblock ``Macroeconomic forecasting through news, emotions and narrative''.
\newblock {\em Expert Systems with Applications\/}~vol. 175, pp. 114760.


\bibitem[\protect\citeauthoryear{{UNHCR}}{{UNHCR}}{2022}]{unhcrfaag}
{UNHCR} 2022, June.
\newblock ``{Figures at a Glance}''.
\newblock \url{https://www.unhcr.org/en-us/figures-at-a-glance.html}.

\bibitem[\protect\citeauthoryear{Vargo, Guo, and Amazeen}{Vargo
  et~al.}{2018}]{vargo2018agenda}
Vargo, C.~J., L.~Guo, and M.~A. Amazeen. 2018.
\newblock ``The agenda-setting power of fake news: A big data analysis of the
  online media landscape from 2014 to 2016''.
\newblock {\em New media \& society\/}~vol. 20 (5), pp. 2028--2049.


\bibitem[\protect\citeauthoryear{Voukelatou, Miliou, Giannotti, and
  Pappalardo}{Voukelatou et~al.}{2022}]{voukelatou2022understanding}
Voukelatou, V., I.~Miliou, F.~Giannotti, and L.~Pappalardo. 2022.
\newblock ``Understanding peace through the world news''.
\newblock {\em EPJ data science\/}~vol. 11 (1), pp. 2.


\bibitem[\protect\citeauthoryear{Yesilbas, Padilla, and Frydenlund}{Yesilbas
  et~al.}{2021}]{yesilbas2021analysis}
Yesilbas, V., J.~J. Padilla, and E.~Frydenlund. 2021.
\newblock ``An analysis of global news coverage of refugees using a big data
  Approach''.
\newblock In {\em Social, Cultural, and Behavioral Modeling: 14th International
  Conference, SBP-BRiMS 2021, Virtual Event, July 6--9, 2021, Proceedings 14},
  pp.  111--120.
\newblock Springer.

\bibitem[\protect\citeauthoryear{Yonamine}{Yonamine}{2013}]{yonamine2013nuanced}
Yonamine, J.~E. 2013.
\newblock {\em A nuanced study of political conflict using the Global Datasets
  of Events Location and Tone (GDELT) dataset}.
\newblock The Pennsylvania State University.


\end{thebibliography}

\section*{Author Biographies}
\label{sec:authorbio}

\textbf{\uppercase{Himarsha R. Jayanetti}} is a Ph.D. student at Old Dominion University working under the supervision of Dr. Michele C. Weigle. She is also a member of the Web Science and Digital Libraries research group as a graduate research assistant. Her research interests are in digital preservation, digital libraries, web science, and social media. Her email address is \email{hjaya002@odu.edu}. 

\textbf{\uppercase{Erika Frydenlund }} is a Research Assistant Professor at VMASC at Old Dominion University. Her research focuses on social justice and mobility, specifically refugees and other forced migrants. She also teaches graduate courses on quantitative and qualitative research methods and refugee studies. Her email address is \email{efrydenl@odu.edu}.

\textbf{\uppercase{Michele C. Weigle}} is a Professor of Computer Science at Old Dominion University. Her research interests include web science, social media, web archiving, and information visualization. Her email address is \email{mweigle@cs.odu.edu}.

\end{document}